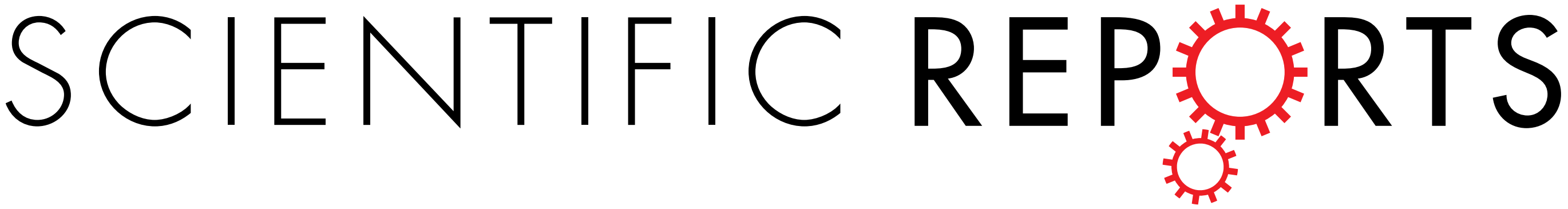

OPEN

# Native point defects of semiconducting layered $Bi_2O_2Se$

Huanglong Li[1], Xintong Xu[2], Yi Zhang[3], Roland Gillen[4], Luping Shi[1] & John Robertson[5]

**$Bi_2O_2Se$ is an emerging semiconducting, air-stable layered material (Nat. Nanotechnol. 2017, *12*, *530*; Nano Lett. 2017, *17*, *3021*), potentially exceeding $MoS_2$ and phosphorene in electron mobility and rivalling typical Van der Waals stacked layered materials in the next-generation high-speed and low-power electronics. Holding the promise of functional versatility, it is arousing rapidly growing interest from various disciplines, including optoelectronics, thermoelectronics and piezoelectronics. In this work, we comprehensively study the electrical properties of the native point defects in $Bi_2O_2Se$, as an essential step toward understanding the fundamentals of this material. The defect landscapes dependent on both Fermi energy and the chemical potentials of atomic constituents are investigated. Along with the bulk defect analysis, a complementary inspection of the surface properties, within the simple context of charge neutrality level model, elucidates the observed n-type characteristics of $Bi_2O_2Se$ based FETs. This work provides important guide to engineer the defects of $Bi_2O_2Se$ for desired properties, which is key to the successful application of this emerging layered material[27].**

The richness of exotic physical properties of layered materials, such as graphene[1–3], transition-metal dichalcogenides[4,5], black phosphorous[6], indium selenide[7], has given rise to diverse intriguing applications, including electronic logic, memory devices[8-10] and optoelectronic devices[11–13]. The extraordinariness of layered materials has also captivated the broad spintronics[2] and piezoelectronics[14] communities. Apart from the unprecedented device functionalities, another natural reason to adopt layered materials is the room for device scaling to low dimension. Hitherto, however, no single layered material stands out as being desirable in all technological aspects. For example, graphene is gapless and therefore unsuitable for conventional transistors, black phosphorous is not air-stable. Thus, immense interest has been sparked in exploring a wider range of layered materials.

Recently, semiconducting $Bi_2O_2Se$ layered material has been synthesized, exhibiting thickness-dependent band gap and air-stability[15,16]. What's more, it is anticipated to have lower in-plane electron effective mass than those of $MoS_2$ and black phosphorous, which has been evidenced by the ultrahigh electron Hall mobility and quantum oscillation at low temperature[15]. Field-effect-transistors (FETs) based on $Bi_2O_2Se$ show encouraging performance and substantial room for further optimization[15], making $Bi_2O_2Se$ a promising candidate for future high-speed and low-power electronic applications. The combination of its exceptional optical and electrical properties has also been exploited in integrated photodetectors of high photoresponsitivity at selective wavelength, holding the promise for next-generation optoelectronic systems[17]. Preceding the showcase in modern electronics, $Bi_2O_2Se$ was a traditional thermoelectric material[18]. The keen interest in the functional versatility of $Bi_2O_2Se$ has been propelled by the theoretical prediction of its larger piezoelectricity and ferroelectricity, under in-plane strain, than those of monolayer $MoS_2$[19], which opens up prospects for energy conversion devices, sensors and non-volatile ferroelectric memories.

Although $Bi_2O_2Se$ is gaining increasing interest from various disciplines, understanding the fundamentals of $Bi_2O_2Se$ as a semiconductor, such as the native point defects, is still premature. Knowing the behaviors of native point defects is essential to the successful application of any semiconductors. These defects, which are naturally present in certain amount, control directly or indirectly the electrical and optical properties of the materials, such as the electrical conductivity. The defect concentration is governed primarily by defect formation energies, especially during the prolonged anneals at elevated temperatures. Thus the formation energies of individual native defects are quantities of central importance. In this letter, we comprehensively investigate the native point defects of $Bi_2O_2Se$ by first-principles calculations. The results provide an important guide to engineer $Bi_2O_2Se$ for desired properties and design functional $Bi_2O_2Se$ devices.

[1]Department of Precision Instrument, Center for Brain Inspired Computing Research, Tsinghua University, Beijing, China. [2]School of Aerospace Engineering, Tsinghua University, Beijing, China. [3]Department of Electronic Engineering, Tsinghua University, Beijing, China. [4]Institute of Physics, Friedrich-Alexander-University of Erlangen-Nürnberg, Nürnberg, Germany. [5]Engineering Department, University of Cambridge, Cambridge, UK. Huanglong Li, Xintong Xu and Yi Zhang contributed equally to this work. Correspondence and requests for materials should be addressed to H.L. (email: li_huanglong@mail.tsinghua.edu.cn)

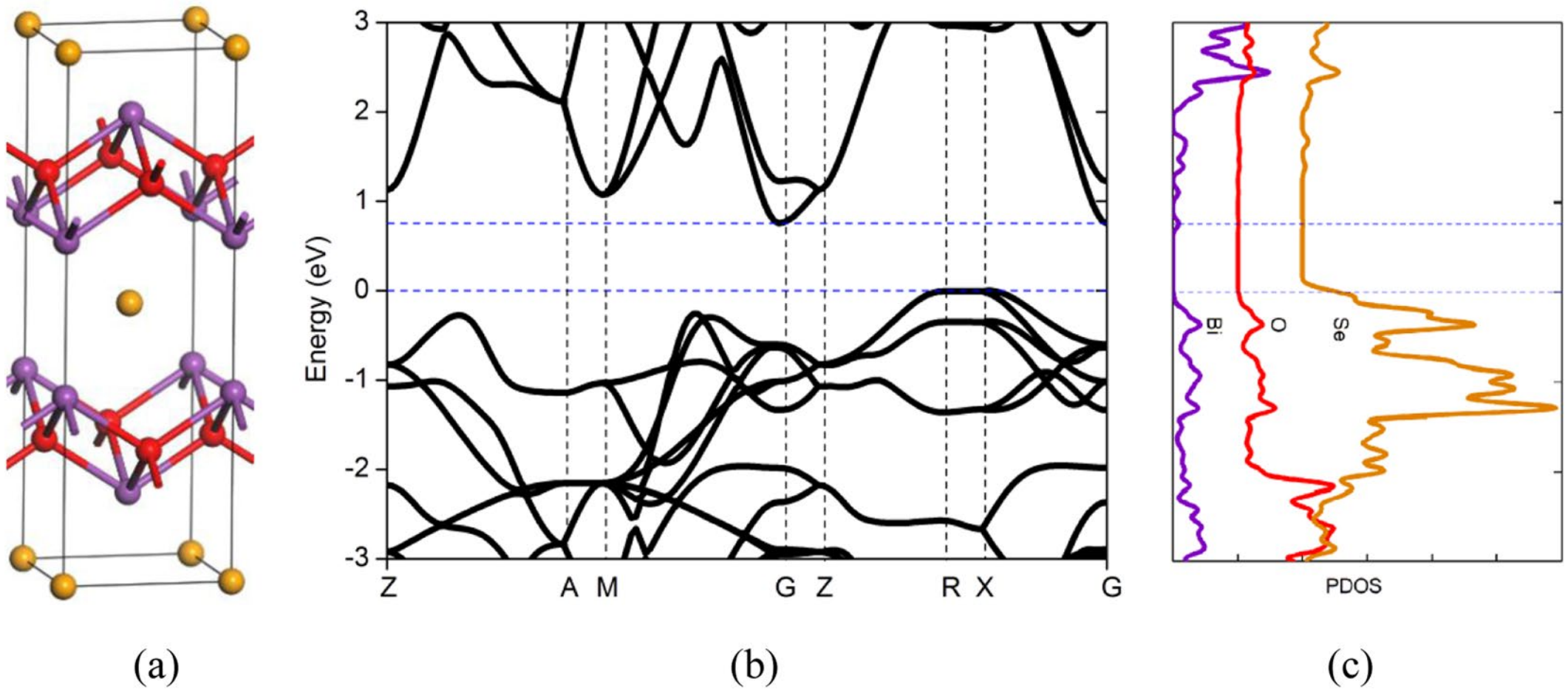

**Figure 1.** (**a**) Atomic structure of $Bi_2O_2Se$, (**b**) band structure of $Bi_2O_2Se$, (**c**) PDOSs of $Bi_2O_2Se$.

## Results and Discussion

Our calculations are based on density functional theory within the generalized gradient approximation,[20] using the Cambridge Sequential Total Energy Package[21]. 90 atoms' $Bi_2O_2Se$ supercell is used as the host of various native point defects, where the lattice constants are fixed to the calculated values. Cutoff energy of the plane wave basis set is 680 eV. All atoms are relaxed in each optimization cycle until atomic forces on each atom are smaller than 0.01 eV $Å^{-1}$ and the energy variation between subsequent iterations falls below $5 \times 10^{-6}$ eV. Total energies are evaluated on $3 \times 3 \times 3$ Monkhorst−Pack k-meshes. Unlike other layered materials which have individual atomic layers stacked by van der Waals interactions, $Bi_2O_2Se$ lacks a well-defined van der Waals gap but displays out-of-plane electrostatic interactions between planar covalently bonded oxide layer ($Bi_2O_2$) and Se square array, as shown in Fig. 1a. The calculated band structure and atomic projected density of states (PDOSs) are shown in Fig. 1b. Indirect band gap of 0.76 eV with conduction band minimum (CBM) near Γ point is in good agreement with the value of 0.80 eV measured by angle-resolved photoemission spectroscopy[15]. The electronic states near the CBM and the valence band maximum (VBM) originate mainly from the Bi and Se/O p-orbital bands, respectively.

$Bi_2O_2Se$ is a ternary semiconductor with ample defect configurations. We consider ten of them in this work, including vacancies, interstitials and antisites in the relevant charge states. The formation energy $\Delta H_f(\alpha, q)$ of defect $\alpha$ in charge state q depends on the chemical potentials $\mu$ of the atomic constituents as well as the electron chemical potential, namely, Fermi energy $\varepsilon_F$. In $Bi_2O_2Se$,

$$\Delta H_f(\alpha, q) = E(\alpha, q) - E(Bi_2O_2Se) + n_{Bi}\mu_{Bi} + n_O\mu_O + n_{Se}\mu_{Se} + q\varepsilon_F + C_{Freysoldt} \tag{1}$$

where $E(\alpha, q)$ is the total energy of the supercell containing a type $\alpha$ defect and charge q, $E(Bi_2O_2Se)$ is the total energy of the defect free supercell, n's and q are the numbers of the atoms and electrons, respectively, that transferred from the defect free supercell to the reservoirs in forming the defect cell. $C_{Freysoldt}$ is the charge state and cell size correction to the defect-formation energy[22]. According to Freysoldt *et al.*, the correction consists of three contributions: a lattice term, a self-interaction term and a potential alignment term[22]. The lattice term accounts for the electrostatic interaction of the defect charge in the supercell with its array of periodic images in the remaining crystal. We use Gaussian defect charge distribution. The lattice energy includes the self-interaction term of the defect charge with its own potential, which must be removed from the correction term. The potential alignment term allows for a meaningful comparison of the formation energies of different charged defects as the charged defect in the supercell will introduce a constant shift of the electrostatic potential and the valence band maximum compared to the ideal host system. As the dielectric anisotropy in layered material systems could be strong, we generalize the Freysoldt scheme to account for the anisotropy. This is achieved by using the dielectric tensor for the calculation of Coulomb interaction potential in the reciprocal space. The dielectric tensor is computed from density-functional perturbation theory. A correction for filling the CBM and emptying the VBM has also been considered[23]. Freysoldt correction leads to well-converged defect formation energies (Fig. S1). The chemical potentials are allowed to vary over a restricted range determined by equilibrium thermodynamics[24]: $\varepsilon_F$ is bound between the VBM and CBM of $Bi_2O_2Se$, and $\mu$'s are bound by the values that (i) will cause precipitation of solid elemental Bi in the trigonal phase, molecular O and solid elemental Se in the trigonal phase, i.e.,

$$\mu_{Bi} < \mu_{Bi}^{solid}, \quad \mu_O < \mu_O^{molecule}, \quad \mu_{Se} < \mu_{Se}^{solid} \tag{2}$$

(ii) maintain the stable $Bi_2O_2Se$ compound, i.e.,

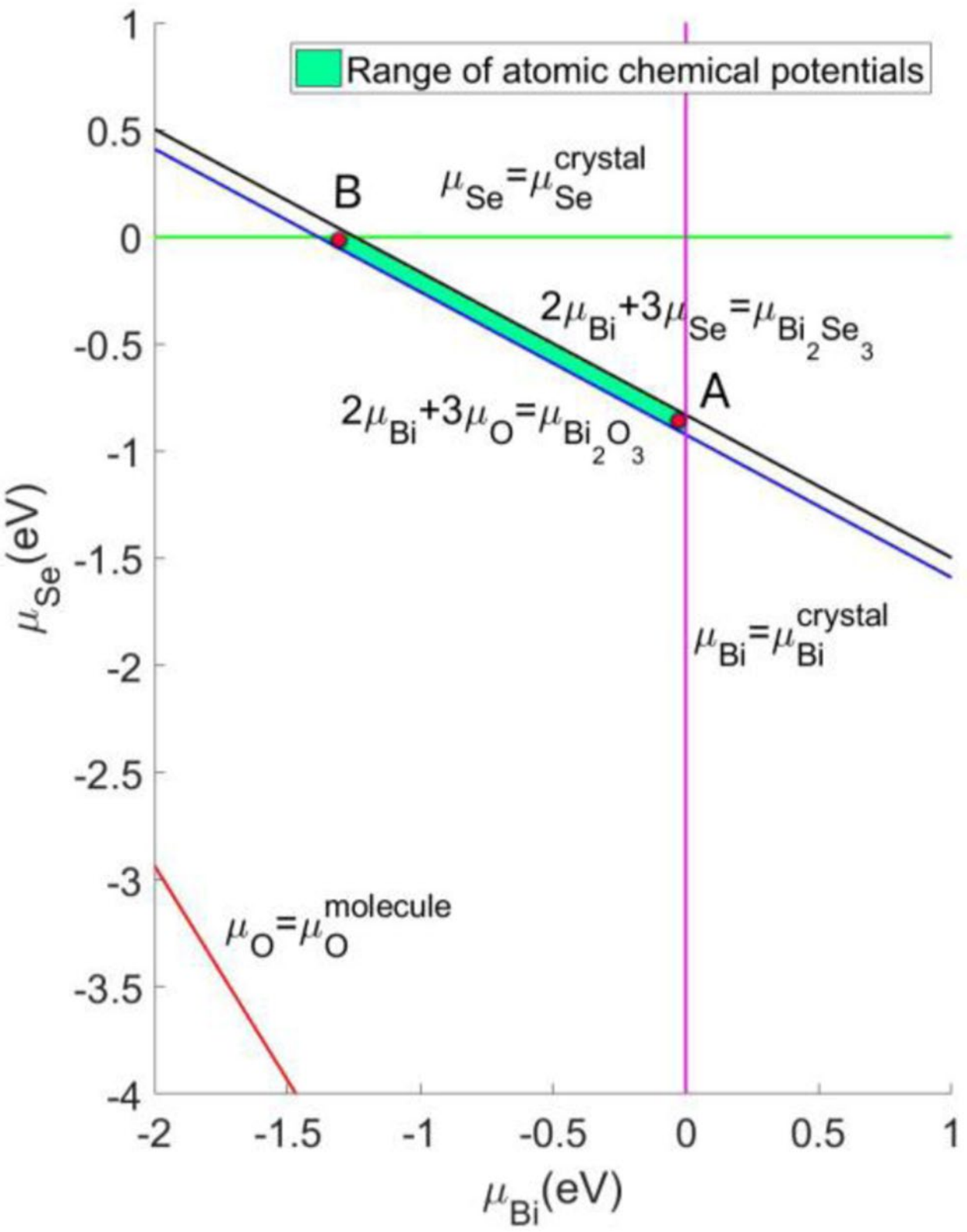

**Figure 2.** Calculated range of atomic chemical potentials for stable $Bi_2O_2Se$. Selected (Se-poor, Bi-rich) and (Se-rich, Bi-poor) conditions are labelled by red point A and B, respectively.

$$2\mu_{Bi} + 2\mu_{O} + \mu_{Se} = \mu_{Bi_2O_2Se} \tag{3}$$

(iii) will cause the formation of solid binaries $Bi_2O_3$ and $Bi_2Se_3$ in the monoclinic and trigonal phases, respectively, i.e.,

$$2\mu_{Bi} + 3\mu_{O} < \mu_{Bi_2O_3}^{solid}, \quad 2\mu_{Bi} + 3\mu_{Se} < \mu_{Bi_2Se_3}^{solid} \tag{4}$$

The calculated range of atomic chemical potentials for stable $Bi_2O_2Se$ is shown on the two-dimensional "$\mu_O$ vs $\mu_{Se}$" plane in Fig. 2. The formation of $Bi_2O_3$ and $Bi_2Se_3$ from their component elements is exothermic by 4.0 eV and 1.2 eV per Bi, respectively. The formation of $Bi_2O_2Se$ from its component binary oxides is exothermic by 0.05 eV per Bi.

We first consider anion deficiency related defects, including O and Se vacancies ($O_v$, $Se_v$), Bi interstitial ($Bi_{in}$) and Bi antisites ($Bi_O$, $Bi_{Se}$). The defect formation energies as a function of Fermi energy $\varepsilon_F$ are shown in Fig. 3, with values of the atomic chemical potentials $\mu_O$ and $\mu_{Se}$ assigned at two representative points on the "$\mu_O$ vs $\mu_{Se}$" plane, respectively. The kinks in the curves for a given defect indicate transitions between different charge states. For both $O_v$ and $Se_v$, they occur exclusively in the positively charged state (charge states from 0 to 2+ are considered) and therefore act as shallow donors. The $\xi(2+/1+)$ transition levels for $O_v$ and $Se_v$ occur at 0.61 eV and 0.64 eV above the VBM, respectively. The atomic structures of $O_v^{1+}$ and $Se_v^{1+}$ are shown in Fig. 4a,b. PDOSs (Fig. S2ab) of $O_v^{\circ}$ and $Se_v^{\circ}$ provide alternative way of understanding the shallow donor effect of $O_v$ and $Se_v$, where we see that electrons are readily provided to the conduction band by thermal excitation at steady state. $Se_v$ remains to be the most stable defect throughout the band gap under the (Se-poor, Bi-rich) condition (Fig. 3a). $O_v$ has slightly higher formation energy. The unintentional n-type doping of $Bi_2O_2Se$ is therefore likely to be due to the existence of $Se_v$ and $O_v$. Under the (Se-rich, Bi-poor) condition, their formation energies increase and they dominate only in the p-type $Bi_2O_2Se$, counteracting the p-type conductivity.

We then consider $Bi_{in}$ at the center of the Se square, which is found to be energetically favorable compared with other interstitial sites. $Bi_{in}$ occurs exclusively in the positively charged state (charge states from 3− to 3+ are considered), thus acting as a shallow donor. A $\xi(3+/2+)$ transition level occurs at 0.19 eV above the VBM. However, because $Bi_{in}$ has quite high formation energy in n-type $Bi_2O_2Se$ even under the (Se-poor, Bi-rich) condition, it is not likely to be present to drive unintentional conductivity of $Bi_2O_2Se$. The atomic structure of $Bi_{in}^{2+}$ is shown in Fig. 5a. It can be seen that the interstitial Bi atom repels the two Bi atoms right below and above in the $Bi_2O_2$ layers to the centers of the O squares.

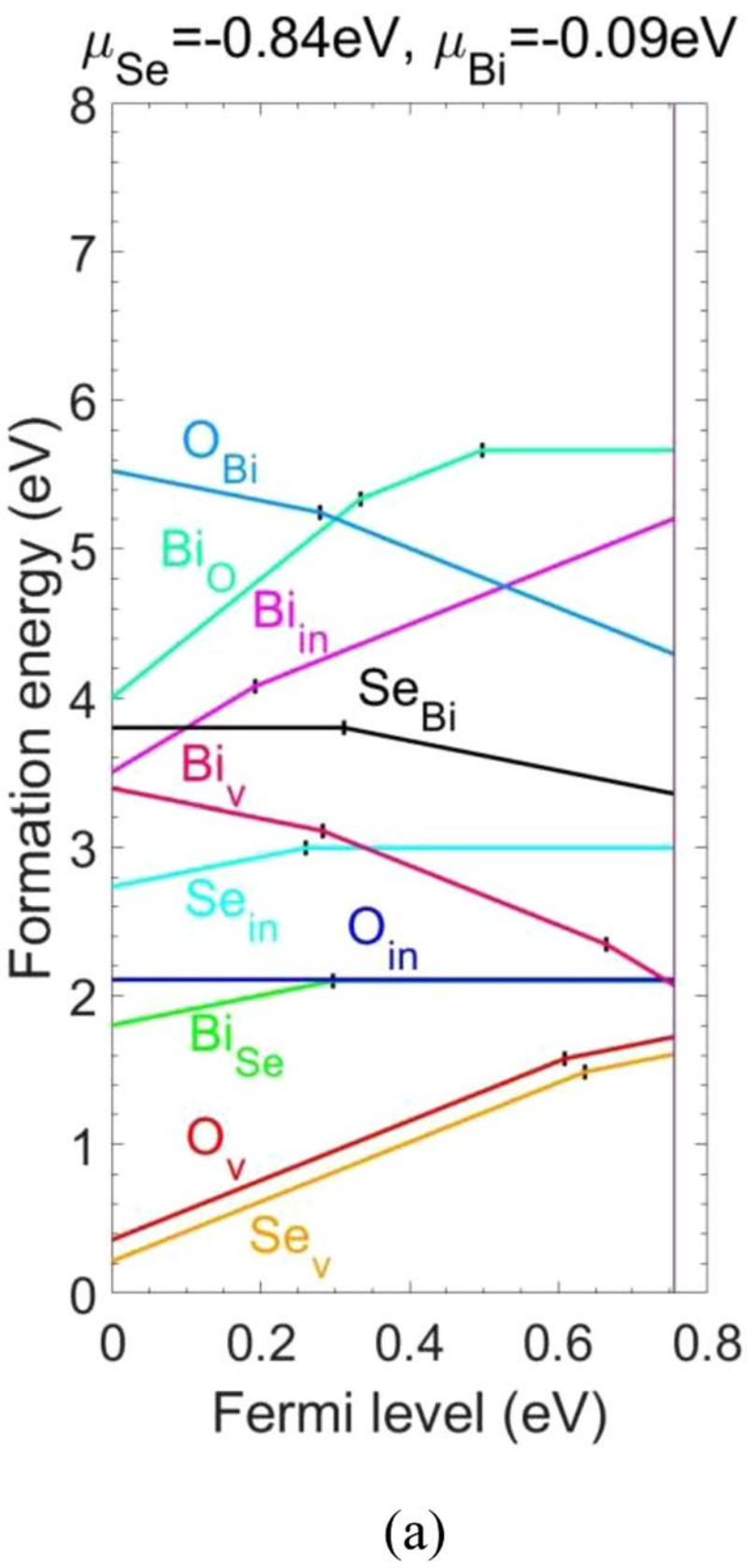

(a)

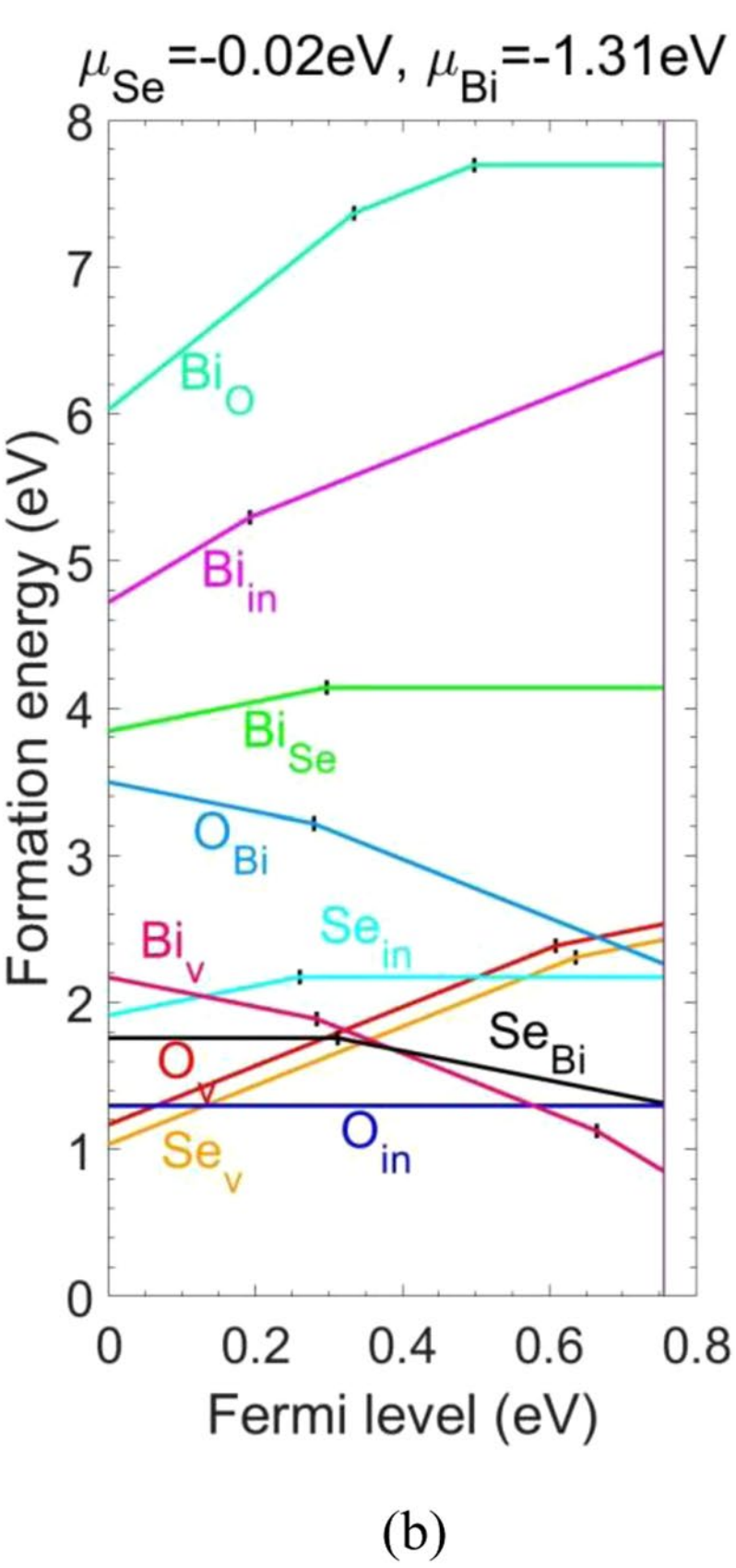

(b)

**Figure 3.** Defect formation energies as a function of Fermi level under the (**a**) (Se-poor, Bi-rich) conditions (cf. point A in Fig. 2) and (**b**) (Se-rich, Bi-poor) ones (cf. point B in Fig. 2). The VBM corresponds to $\varepsilon_F = 0$ eV and the CBM corresponds to $\varepsilon_F = 0.76$ eV which is indicated by a vertical line.

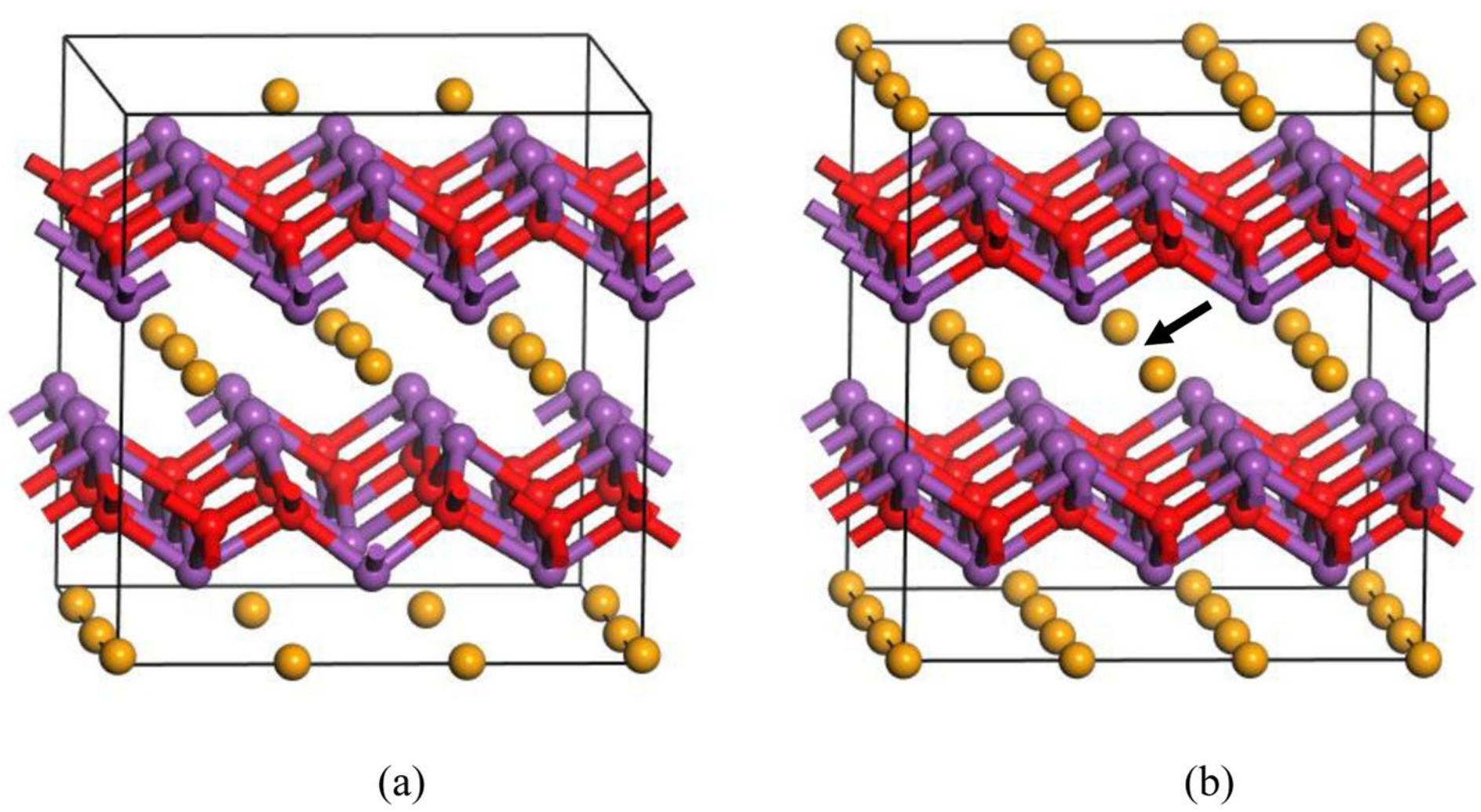

(a) (b)

**Figure 4.** Atomic structures of (**a**) $O_v^{1+}$ and (**b**) $Se_v^{1+}$.

Next, we consider two kinds of Bi antisites. For $Bi_{Se}$, a $\xi(1+/0)$ transition level occurs at 0.30 eV above the VBM (charge states from 1− to 5+ are considered), indicating that it is a deep donor center. It is the third most stable defect next to $Se_v$ and $O_v$ in p-$Bi_2O_2Se$ under the (Se-poor, Bi-rich) condition, compensating the p-type conductivity. The formation energy of $Bi_{Se}$ rapidly increases with anion chemical potentials and $Bi_{Se}$ becomes less likely to exist under the (Se-rich, Bi-poor) condition. The atomic structure of neutral $Bi_{Se}$ is shown in Fig. 5b.

For $Bi_O$, transition levels $\xi(4+/2+)$ and $\xi(2+/0)$ occur at 0.33 eV and 0.50 eV above the VBM, respectively (charge states from 1− to 5+ are considered). Therefore, it is a deep donor center and it compensates the p-type conductivity of $Bi_2O_2Se$. However, it has quite high formation energy even in the (Se-poor, Bi-rich) condition, so it is not likely to exist. It is obvious that $Bi_O$ is a negative-U defect, with transition level $\xi(4+/3+)$ higher than $\xi(3+/2+)$, and $\xi(2+/1+)$ higher than $\xi(1+/0)$. Negative-U behavior has been typically related to unusually large local lattice relaxations that stabilize particular charge states. Here, substitutional Bi in the 4+ charged state undergoes displacement ~1 Å vertical to the $Bi_2O_2$ plane (Fig. 6), whereas it remains almost onsite in the case of $Bi_O^{3+}$ and $Bi_O^{2+}$. At the same time, there is ~0.2 Å difference in the displacement of a neighboring Bi atom vertical to the $Bi_2O_2$ plane between $Bi_O^{\circ}$ and $Bi_O^{1+}$/ $Bi_O^{2+}$, whereas the latter two are comparable (not shown).

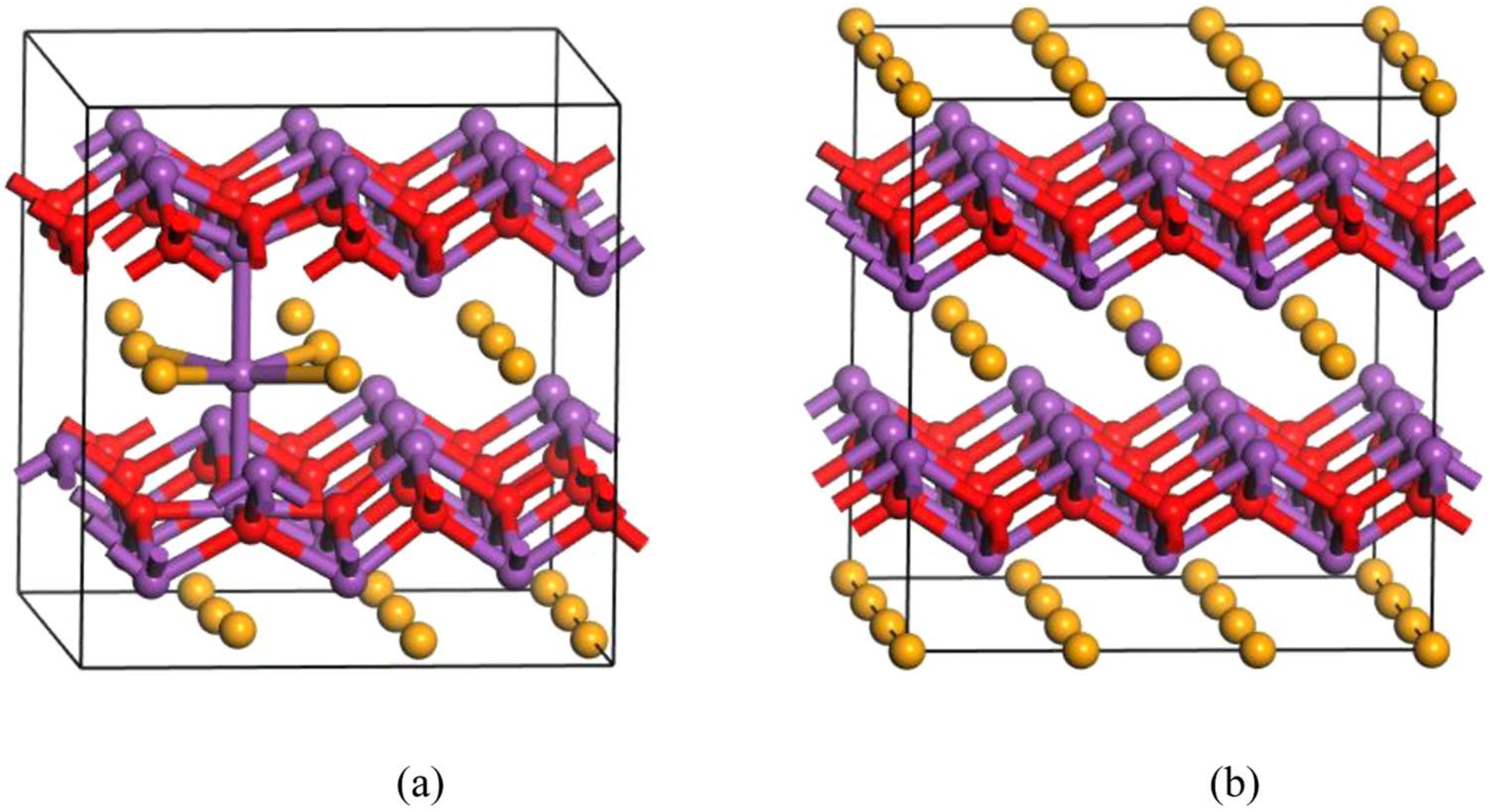

**Figure 5.** Atomic structures of (**a**) $Bi_{in}^{2+}$ and (**b**) neutral $Bi_{Se}$.

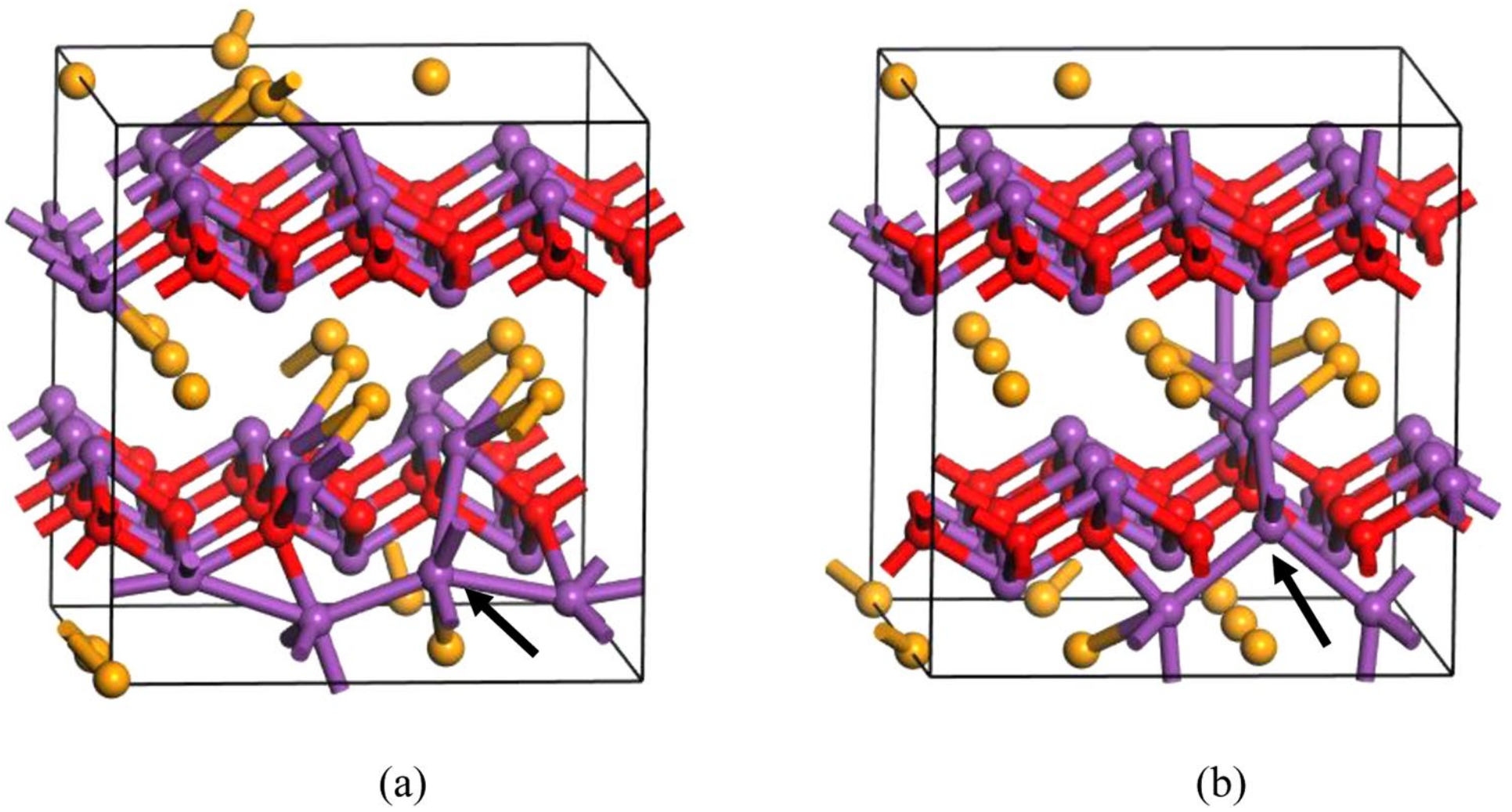

**Figure 6.** Atomic structures of (**a**) $Bi_O^{4+}$ and (**b**) neutral $Bi_O^{2+}$.

Next, we investigate cation deficiency related native point defects, including O and Se interstitials ($O_{in}$, $Se_{in}$), Bi vacancies ($Bi_v$) and O/Se antisites ($O_{Bi}$, $Se_{Bi}$). For $O_{in}$ right above a Se ion, which is found to be energetically favorable, it repels the Se ion out of the Se square plane and forms seleninyl ion $SeO^{-2+q}$ (charge states q from 2− to 2+ are considered). The atomic structure of $O_{in}^{\circ}$ is shown in Fig. 7a. $O_{in}$ remains neutral throughout the band gap, rendering on average −1 intermediate oxidation state for Se and O atoms in the seleninyl ion. Its formation energy is relatively high under the (Se-poor, Bi-rich) condition compared with $O_v$ and $Se_v$, but is reduced under the (Se-rich, Bi-poor) condition, becoming the most favorable defect in the Fermi energy range between 0.13 eV and 0.58 eV above the VBM. For $Se_{in}$ at the edge center of the Se square, which is found to be energetically favorable, it repels the two neighboring Se ions at the vertices, forming triselenide anion $Se_3^{-4+q}$ (charge states q from 2− to 4+ are considered). The atomic structure of $Se_{in}^{\circ}$ is shown in Fig. 7b. $Se_{in}$ assumes positively charged state as long as the Fermi level is below the transition level $\xi(1+/0)$ at 0.26 eV above the VBM, acting as an acceptor compensating center in p-type $Bi_2O_2Se$. On average, each Se atom in the triselenide anion acquires intermediate oxidation state between −2 and 0. It has higher formation energy than $O_{in}$ does. PDOSs of $O_{in}^{\circ}$ ($Se_{in}^{0}$) (Fig. 7c,d) show that the seleninyl ion $SeO^{-2}$ (triselenide anion $Se_3^{-4}$) has filled antibonding frontier orbital states below (above) the VBM. It is well known that chalcogen elements (S, Se, Te) have strong tendency to form polychalcogenide anion $X_N^{-Q}$ (X: chalcogen element) of variable chain length N, with average oxidation state assumed by each atom intermediate between −2 and 0. This rationalizes the favorable nonnegative charge states for $O_{in}$ and $Se_{in}$.

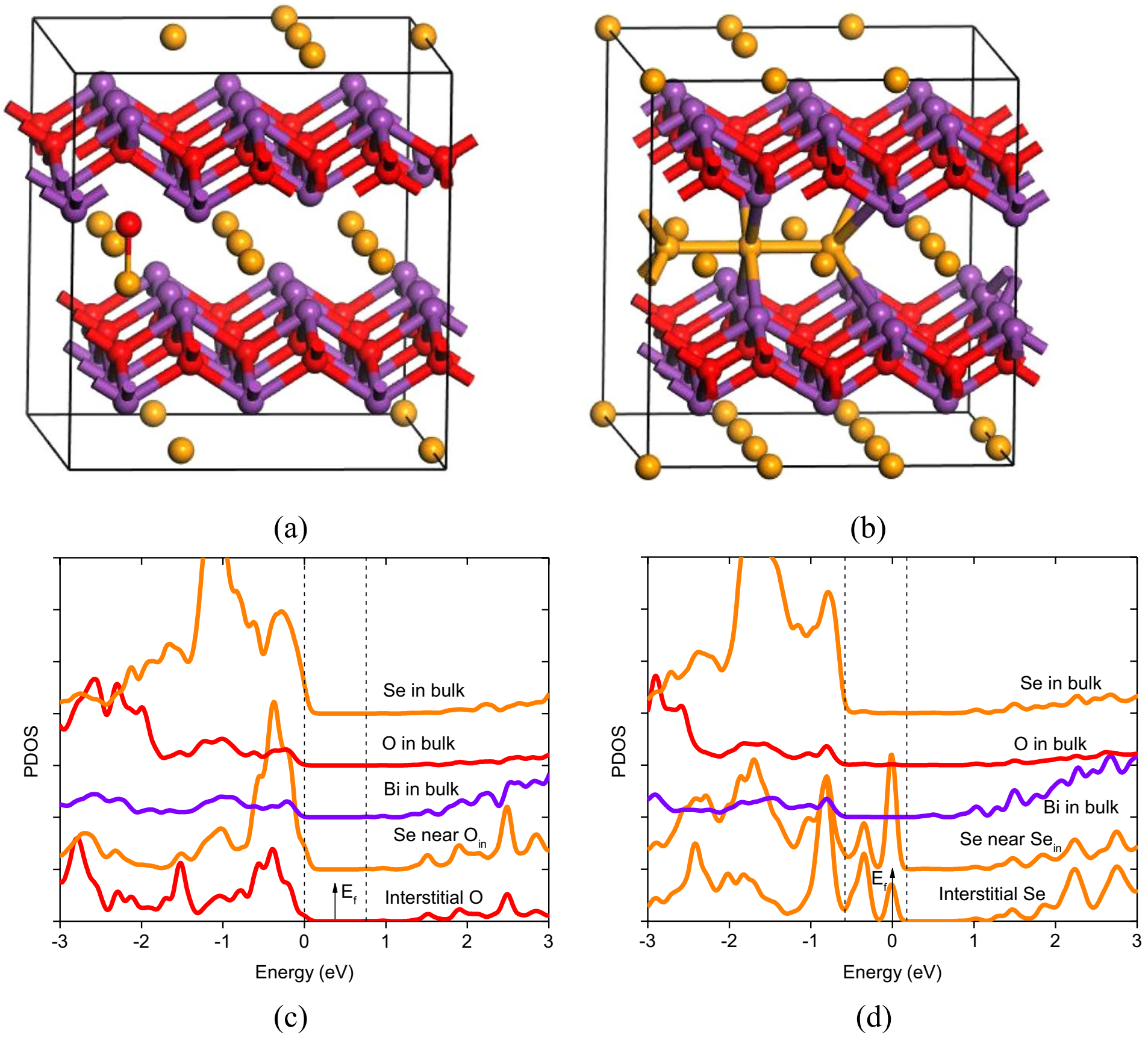

**Figure 7.** Atomic structures of charge neutral (**a**) $O_{in}$ and (**b**) $Se_{in}$, PDOSs of charge neutral (**c**) $O_{in}$ and (**d**) $Se_{in}$. The vertical dash lines indicate the VBM and CBM of the host system, drawn from the PDOSs of the atoms in bulk which serve as references.

Next, we consider $Bi_v$. It is a shallow acceptor which occurs exclusively in the negative charge state, with transition levels $\xi(1-/2-)$ and $\xi(2-/3-)$ occur at 0.28 eV and 0.66 eV above the VBM, respectively (charge states from 3- to 0 are considered). PDOSs (Fig. S3) of $Bi_v^0$ provide alternative way of understanding the shallow acceptor effect of $Bi_v$, where we see that it is ready to accept electrons near the VBM by thermal excitation at steady state, leaving holes in the valence band. The formation energy of $Bi_v$ under the (Se-poor, Bi-rich) condition is relatively high compared with those of $Se_v$ and $O_v$. Under the (Se-rich, Bi-poor) condition, however, the formation energy of $Bi_v$ rapidly decreases and $Bi_v$ becomes the dominant defect in n-type $Bi_2O_2Se$, compensating the prevalent conductivity. The atomic structure of $Bi_v^{3-}$ is shown in Fig. 8a.

Finally, we consider $O_{Bi}$ and $Se_{Bi}$. $O_{Bi}$ occurs exclusively in the negative charge state (charge states from 5− to 1+ are considered) and acts as a shallow acceptor. The transition level $\xi(1-/2-)$ occur at 0.28 eV above the VBM. The atomic structure of $O_{Bi}^{2-}$ is shown in Fig. 8b. The substitutional O atom undergoes large displacement to the Se square layer, locating in the center of two neighboring Se atoms. The formation energy of $O_{Bi}$ is high even under the (Se-rich, Bi-poor) condition. Thus, it is not likely to exist.

For $Se_{Bi}$, the transition level $\xi(0/1-)$ occurs at 0.31 eV above the VBM. Thus, it compensates donors. The formation energy of $Se_{Bi}$ under the (Se-poor, Bi-rich) condition is quite high but rapidly decreases with increasing anion chemical potentials. Under the (Se-rich, Bi-poor) condition, $Se_{Bi}$ becomes the second most likely compensating center next to $Bi_v$ in n-type $Bi_2O_2Se$. The atomic structure of $Se_{Bi}^{1-}$ in Fig. 8c shows that the substitutional Se atom displaces toward the center of the underlying Se square.

Here to, we have studied the electrical properties of ten types of native point defects. As previously pointed out, the electrical conductivity of material can be significantly affected by its native point defects. In $Bi_2O_2Se$ FETs, whose channels remain conducting at $V_g = 0$, the total resistance decreases with increasing gate bias[15], which is a clear signature of n-type characteristics. Figure 3 implies the possibility of (Se-poor, Bi-rich) fabrication condition for the $Bi_2O_2Se$ FETs since $Se_v$ and $O_v$, which are shallow donors, are the most likely defects under this condition.

The conductivity of $Bi_2O_2Se$ can alternatively be understood within the context of charge neutrality level (CNL) model[25]. The CNL model is useful because it is simple and gives good chemical trends, while requiring no specified details of surface chemical bonding which are outside of the scope of this work. The CNL is the demarcation between the surface states that are predominantly donor-like (valence band states) and acceptor-like (conduction band states), namely, at CNL they have equal densities. Mathematically, the CNL is the branch point of the imaginary bulk band structure of the semiconductor. It is calculated as the zero of the Greens function of the band structure averaged over the Brillouin zone:

$$G(E) = \int_{BZ} \frac{N(E')dE'}{(E - E') + i\delta} = 0 \quad (5)$$

where $\delta$ is a small number to be used if the CNL lies inside a band. It can also be expressed as a sum over special points of the Brillouin zone (such as the Monkhorst–Pack grid)[26]:

$$G(E) = \sum_n \frac{1}{E - E_n} \quad (6)$$

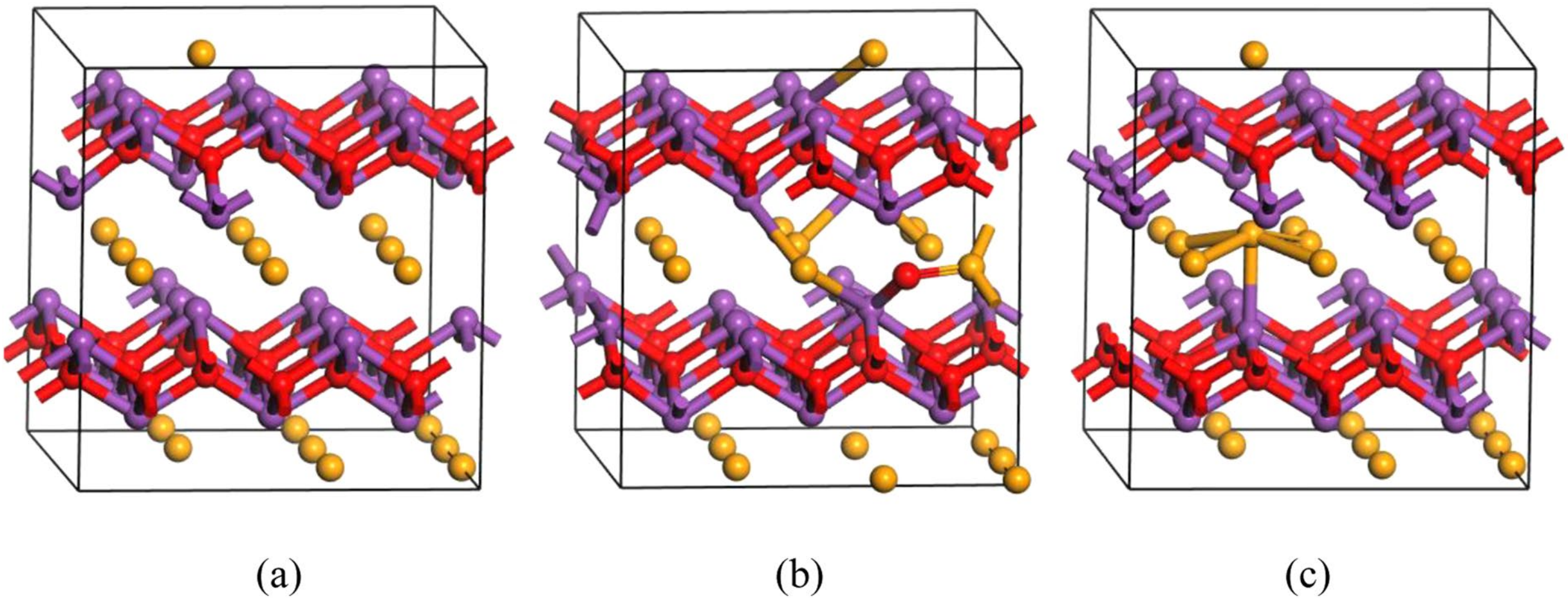

**Figure 8.** Atomic structure of (**a**) $Bi_v^{3-}$, (**b**) $O_{Bi}^{2-}$ and (**c**) $Se_{Bi}^{1-}$.

The CNL is then a weighted average of the valence and conduction band DOS:

$$E_{CNL} = \frac{N_v E_c + N_c E_v}{N_v + N_c} \quad (7)$$

In this definition, the CNL is an intrinsic property of the bulk semiconductor; it does not depend on the interface, or interface bonding, or whatever it is attached to.

According to equation (7), the CNL of $Bi_2O_2Se$ is calculated to be ~0.7 eV above the VBM, very close to the CBM. In the absence of gate bias, the Fermi level at the surface of $Bi_2O_2Se$ is aligned with the CNL to ensure charge neutrality, resulting in surface electron accumulation. This explains why $Bi_2O_2Se$ based FETs are in the ON state when $V_g = 0$ and show n-type behavior.

## Conclusion

In summary, we have systematically studied all anion and cation deficiency related native point defects of $Bi_2O_2Se$ in all relevant charge states. The abounding defect behaviors resulting from the ternary elemental compositions and the unique stacking structure are analyzed. Defect landscape is found to vary with Fermi energy and the chemical potentials of the atomic constituents. Our results suggest the possibility of (Se-poor, Bi-rich) fabrication condition of the previously reported $Bi_2O_2Se$ FETs[15]. Under this condition, $Se_v$ and $O_v$ are the dominant defects and they act as shallow donors, accounting for the unintentional n-type conductivity of $Bi_2O_2Se$. Alternatively, the n-type characteristics of $Bi_2O_2Se$ FETs can also be understood in the context of the CNL model. The CNL of $Bi_2O_2Se$ is computed to be close to the CBM, resulting in surface electron accumulation. This work provides important guide to engineer the defects of $Bi_2O_2Se$ for desired properties, which is key to the successful application of this emerging layered material.

## Acknowledgements
H. Li thanks Beijing Natural Science Foundation (No. 4164087) and National Natural Science Foundation (No. 61704096) for financial support. R. Gillen thanks Deutsche Forschungsgemeinschaft (DFG) within the Cluster of Excellence Engineering of Advanced Materials (project EXC 315) (Bridge Funding) for financial support. L.P. Shi thanks National Natural Science foundation (No. 61603209, 61475080, 61327902), SuZhou-Tsinghua innovation leading program (2016SZ0102) and Beijing Innovation Centre for Future Chip for financial support. J. Robertson thanks EPSRC for financial support. Computational resources are provided by high performance computing service of Tsinghua National Laboratory for Information Science and Technology and high performance computing service of University of Cambridge.

## Author Contributions
H. Li conceived this work. H. Li, X.X and Y. Zhang performed DFT calculations. H. Li wrote the manuscript. All authors discussed the results and implications and commented on the manuscript at all stages.

## Additional Information
**Supplementary information** accompanies this paper at https://doi.org/10.1038/s41598-018-29385-8.

**Competing Interests:** The authors declare no competing interests.

**Publisher's note:** Springer Nature remains neutral with regard to jurisdictional claims in published maps and institutional affiliations.